\documentclass[preprint,amsfonts,amssymb,amsmath,%
tightenlines,nofootinbib,showpacs,showkeys]{revtex4}

\newcommand{\cN}{\mathcal{N}}
\newcommand{\cV}{\mathcal{V}}
\newcommand{\cW}{\mathcal{W}}
\newcommand{\bA}{\boldsymbol{A}}
\newcommand{\bB}{\boldsymbol{B}}
\newcommand{\bC}{\boldsymbol{C}}

\newcommand{\bbR}{\mathbb{R}}
\newcommand{\rnu}{\sqrt{\nu}}
\newcommand{\Lsl}{\mathfrak{sl}}
\newcommand{\Int}{\int\!\!}

\begin{document}


\title{A Family of Quasi-solvable Quantum Many-body Systems}

\author{Toshiaki Tanaka}
\email{totanaka@yukawa.kyoto-u.ac.jp}
\affiliation{Faculty of Integrated Human Studies,\\
 Kyoto University, Kyoto 606-8501, Japan}
\altaffiliation{Present address: Yukawa Institute for
Theoretical Physics, Kyoto University,
Kyoto 606-8502, Japan}



\begin{abstract}
We construct a family of quasi-solvable quantum many-body systems
by an algebraic method. The models contain up to two-body interactions
and have permutation symmetry. We classify these models under the
consideration of invariance property. It turns out that this family
includes the rational, hyperbolic (trigonometric) and elliptic
Inozemtsev models as particular cases.
\end{abstract}



\preprint{KUCP-0202}

\maketitle


\section{\label{sec:intro}Introduction}

New findings of solvable or integrable models have stimulated
development of new and wide research directions and ideas
in both physics and mathematics.
The discovery of \textit{quasi-solvability} in quantum
mechanics~\cite{TuUs1} is a typical example.
By quasi-solvability we mean that a part of the spectra can be
solved, at least, algebraically\footnote{The term \textit{
quasi-exact solvability} has been widely used in this meaning.
However, we keep it to express the case where the state vectors
corresponding to the solvable spectra are normalizable. Importance
of this distinction is explained in Refs.~\cite{AST2,ST1}}.
One of the most successful approach to construct a quasi-solvable
model is the algebraic method
introduced by Turbiner in 1988~\cite{Turbi1}, in which
a family of quasi-solvable one-body models was constructed by
the $\Lsl(2)$ generators on a polynomial space. This family was
later completely classified under the consideration of the
$GL(2,\bbR)$ invariance of the models~\cite{LoKaOl3,LoKaOl4}.
Recently, this family have been paid much attention to in the context
of $\cN$-fold supersymmetry~\cite{AKOSW2,ASTY,AST1,KlPl2,KlPl3,%
DoDuTa1,DoDuTa2,AST2,ANST1,ANST2,ST1,Tanak2}.
Several attempts were made to construct quasi-solvable many-body
models by naive extension to higher-rank algebras. Especially,
construction of two-body problems by the rank 2 algebras was
extensively investigated~\cite{LoKaOl4,ShTu1,Shifm1,LoKaOl1,%
LoKaOl2,Turbi2,Ushve}.
These approaches however led to Schr\"{o}dinger operators
in curved space in general and could hardly apply to $M$-body
($M>2$) problems.

In 1995, a significant progress was made in Ref.~\cite{RuTu1},
where the exact solvability of the rational
and trigonometric $A$ type Calogero--Sutherland (CS)
models~\cite{Calog1,Suthe1,Moser1} for any finite number of
particles were shown by a similar algebraic method. The key
ingredient is the introduction of the elementary symmetric
polynomials which reflect the permutation symmetry
of the original models. The algebra for the $M$-body system
is $\Lsl(M+1)$. This idea was further applied to
show the exact solvability of the rational and trigonometric
$A$ and $BC$ type CS models and their supersymmetric
generalizations~\cite{BrTuWy1}, and to show the quasi-exact
solvability of various deformed CS models~\cite{MiRoTu1,HoSh1}.
Therefore, one can say the approach starting from Ref.~\cite{RuTu1}
is, up to now, the most successful in investigating quasi-solvable
quantum many-body problems. However, one has not yet known all the
models that can be obtained by this approach. In other words, we
have not obtained the classification of these $\Lsl(M+1)$ $M$-body
models like that of the $\Lsl(2)$ one-body models. Recently, this
classification problem was partly accessed in Ref.~\cite{UlLoRo1}
though, as was stressed by the authors themselves, the results depend
on the specific ansatz and thus are incomplete.
In this Letter, we will show the complete classification of the
quantum many-body systems with \textit{up to two-body interactions}
which can be constructed by the $\Lsl(M+1)$ method.

\section{\label{sec:const}Construction of the Models}

Consider an $M$-body quantum Hamiltonian,
\begin{eqnarray}
H_{\cN}=-\frac{1}{2}\sum_{i=1}^{M}\frac{\partial^{2}}%
{\partial q_{i}^{2}}+V(q_{1},\ldots,q_{M}),
\label{eqn:psham}
\end{eqnarray}
which possesses permutation symmetry, that is,
\begin{eqnarray}
V({}\ldots,q_{i},\ldots,q_{j},\ldots)=
V({}\ldots,q_{j},\ldots,q_{i},\ldots),
\label{eqn:pspot}
\end{eqnarray}
for $\forall\, i\neq j$.
To algebraize the Hamiltonian (\ref{eqn:psham}), we will proceed
the following three steps. At first, we make a \textit{gauge}
transformation on the Hamiltonian (\ref{eqn:psham}):
\begin{eqnarray}
\tilde{H}_{\cN}=e^{\cW(q)}H_{\cN}e^{-\cW(q)}.
\label{eqn:gtham}
\end{eqnarray}
The function $\cW(q)$ is to be determined later and plays the role
of the superpotential when the system Eq.~(\ref{eqn:psham}) is
supersymmetric. As in Eq.~(\ref{eqn:gtham}), we will hereafter
attach tildes to both operators and vector spaces to indicate
that they are quantities gauge-transformed from the original ones.
In the next,
we change the variables $q_{i}$ to $h_{i}$ by a function
$h$ of a single variable; $h_{i}=h(q_{i})$.
Note that the way of changing of the variables preserves
the permutation symmetry. The third step is the introduction of
elementary symmetric polynomials of $h_{i}$ defined by,
\begin{eqnarray}
\sigma_{k}(h)=\sum_{i_{1}<\cdots <i_{k}}h_{i_{1}}\cdots h_{i_{k}}
\quad (k=1,\ldots,M),
\label{eqn:sigma}
\end{eqnarray}
from which we further change the variables to $\sigma_{i}$.
Then, we choose a set of components of the $\cN$-fold supercharges in
terms of the above variables $\sigma_{i}$ as follows:
\begin{eqnarray}
\tilde{P}_{\cN}^{\{i\}}=\frac{\partial^{\cN}}%
{\partial\sigma_{i_{1}}\cdots\partial\sigma_{i_{\cN}}}
\quad (1\le i_{1}\le\cdots\le i_{\cN}\le M),
\label{eqn:cptsc}
\end{eqnarray}
where $\{i\}$ is an abbreviation of the set $\{i_{1},\ldots,i_{\cN}\}$.
Using these $\cN$-fold supercharges, we define the vector space
$\tilde{\cV}_{\cN}\equiv\bigcap_{\{i\}}\ker \tilde{P}_{\cN}^{\{i\}}$
, which now becomes,
\begin{eqnarray}
\tilde{\cV}_{\cN}
=\text{span }\left\{\sigma_{1}^{n_{1}}\cdots\sigma_{M}^{n_{M}}:
0\le\textstyle{\sum_{i=1}^{M}}n_{i}\le\cN -1\right\}.
\label{eqn:svspc}
\end{eqnarray}
For given $M$ and $\cN$, the dimension of the vector space
(\ref{eqn:svspc}) becomes,
\begin{eqnarray}
\dim \tilde{\cV}_{\cN}=\sum_{n=0}^{\cN -1}\frac{(n+M-1)!}{n!\, (M-1)!}
=\frac{(\cN +M-1)!}{(\cN -1)!\, M!}.
\label{eqn:dimvs}
\end{eqnarray}
We will construct the system (\ref{eqn:gtham}) to be quasi-solvable
so that the solvable subspace is given by just
Eq.~(\ref{eqn:svspc}). This can be achieved by imposing the
following quasi-solvability condition~\cite{AST2,ANST1,Tanak2},
\begin{eqnarray}
\tilde{P}_{\cN}^{\{i\}}\tilde{H}_{\cN}\tilde{\cV}_{\cN}=0
\quad\text{for}\quad \forall\{i\}.
\label{eqn:qscon}
\end{eqnarray}
The general solution of Eq.~(\ref{eqn:qscon}) can be obtained in
completely the same way as shown in Refs.~\cite{ANST1,Tanak2}.
As in the case of the one-body models, it is sufficient to
find differential operators up to the second-order as
solutions for $\tilde{H}_{\cN}$ since we are constructing
a Schr\"{o}dinger operator in the original variables $q_{i}$.
It turns out that the general solution which contains up to the
second derivatives takes the following form,
\begin{eqnarray}
\tilde{H}_{\cN}=-
\!\!\!
\sum_{\kappa,\lambda,\mu,\nu=0}^{M}
\!\!\!
A_{\kappa\lambda,\mu\nu}E_{\kappa\lambda}E_{\mu\nu}
+\sum_{\kappa,\lambda=0}^{M}B_{\kappa\lambda}E_{\kappa\lambda}
-C,
\label{eqn:gsham}
\end{eqnarray}
where $A_{\kappa\lambda,\mu\nu}$, $B_{\kappa\lambda}$, $C$ are
arbitrary constants, and $E_{\kappa\lambda}$ are the first-order
differential operators which constitute the Lie algebra $\Lsl(M+1)$:
\begin{subequations}
\label{eqns:gensl}
\begin{eqnarray}
E_{0i}&=&\frac{\partial}{\partial\sigma_{i}},\quad
E_{ij}=\sigma_{i}\frac{\partial}{\partial\sigma_{j}},
\label{eqn:gensl1}\\
E_{i0}&=&\sigma_{i} E_{00}=\sigma_{i}\left(\cN -1-\sum_{k=1}^{M}
\sigma_{k}\frac{\partial}{\partial\sigma_{k}}\right).
\label{eqn:gensl2}
\end{eqnarray}
\end{subequations}
If we explicitly express the general solution (\ref{eqn:gsham})
in terms of $\sigma_{i}$, we obtain the following expression,
\begin{eqnarray}
\tilde{H}_{\cN}&=&-\sum_{k,l=1}^{M}\bigl[\bA_{0}(\sigma)
\sigma_{k}\sigma_{l}-\bA_{k}(\sigma)\sigma_{l}+\bA_{kl}(\sigma)\bigr]
\frac{\partial^{2}}{\partial\sigma_{k}\partial\sigma_{l}}\nonumber\\
&&{}+\sum_{k=1}^{M}\bigl[\bB_{0}(\sigma)\sigma_{k}-\bB_{k}(\sigma)\bigr]
\frac{\partial}{\partial\sigma_{k}}-\bC(\sigma),
\label{eqn:sgham}
\end{eqnarray}
where $\bA_{\kappa}$, $\bA_{kl}$, $\bB_{\kappa}$ and $\bC$ are
second-degree polynomials of several variables.

One of the most difficult problems one would come across
in the algebraic approach to the quasi-solvable \textit{quantum}
many-body systems
is to solve the canonical-form condition:
\begin{eqnarray}
H_{\cN}=e^{-\cW(q)}\tilde{H}_{\cN}e^{\cW(q)}=-\frac{1}{2}
\sum_{i=1}^{M}\frac{\partial^{2}}{\partial q_{i}^{2}}+V(q).
\label{eqn:cfcon}
\end{eqnarray}
If the Hamiltonian (\ref{eqn:sgham}) is gauge-transformed
back to the original one, it in general does not take the canonical
form of the Schr\"odinger operator like Eq.~(\ref{eqn:psham})
and one can hardly solve, for arbitrary $M$, the conditions
under which a gauge-transform of Eq.~(\ref{eqn:sgham})
could be cast in the Schr\"odinger form. This difficulty can,
however, be partly overcome by the following observation.
Suppose we can solve the canonical-form
condition for an $M$ and obtain a quasi-solvable $M$-body Hamiltonian
constructed from the $\Lsl(M+1)$ generators, which would have
the following form:
\begin{eqnarray}
H_{\cN}=-\frac{1}{2}\sum_{i=1}^{M}\frac{\partial^{2}}{%
 \partial q_{i}^{2}}+\sum_{i=1}^{M}V_{1}(q_{i})+\sum_{i<j}^{M}
 V_{2}(q_{i},q_{j})+\dots +V_{M}(q_{1},\dots,q_{M}).
\label{eqn:mbqsh}
\end{eqnarray}
Then, we can get a quasi-solvable $M$-body model with up to
the two-body interactions if we turn off all the coupling constants
of the interactions except for the one- and two-body ones:
\begin{eqnarray}
H_{\cN}=-\frac{1}{2}\sum_{i=1}^{M}\frac{\partial^{2}}{%
 \partial q_{i}^{2}}+\sum_{i=1}^{M}V_{1}(q_{i})+\sum_{i<j}^{M}
 V_{2}(q_{i},q_{j}).
\label{eqn:2bqsh}
\end{eqnarray}
The resultant model (\ref{eqn:2bqsh}) should be, when we put $M=2$,
identical with one of
the two-body models constructed from the $\Lsl(3)$ generators.
This comes from the fact that the gauged Hamiltonian (\ref{eqn:sgham})
constructed from the $\Lsl(M+1)$ generators reduces to the one
constructed from the $\Lsl(3)$ generators if we put
$M=2$ and $h_{i}=0$ for $i>2$.

Therefore, as far as up to two-body interactions are concerned,
it is sufficient to solve the $M=2$ case by virtue
of the permutation symmetric construction. 
We have found that we can actually solve the canonical-form
condition for $M=2$ and that $\tilde{H}_{\cN}$ for $M\ge 2$ must have
the following expression in terms of the variables $h_{i}$,
\begin{eqnarray}
\tilde{H}_{\cN}(h)&=&-\sum_{i=1}^{M}P(h_{i})\frac{\partial^{2}}%
{\partial h_{i}^{2}}\nonumber\\
&&{}-\sum_{i=1}^{M}\left[Q(h_{i})-\frac{\cN-2+(M-1)c}{2}P'(h_{i})\right]
\frac{\partial}{\partial h_{i}}\nonumber\\
&&{}-2c\sum_{i\neq j}^{M}\frac{P(h_{i})}{h_{i}-h_{j}}\frac{\partial}%
{\partial h_{i}}-\bC\bigl(\sigma(h)\bigr),
\label{eqn:htham}
\end{eqnarray}
where $\bC$ is given by,
\begin{eqnarray}
\bC\bigl(\sigma(h)\bigr)&=&
\frac{\cN -1}{12}\bigl[\cN -2+2(M-1)c\bigr]\sum_{i=1}^{M}P''(h_{i})
\nonumber\\
&&{}-\frac{\cN -1}{2}\sum_{i=1}^{M}Q'(h_{i})
-\frac{\cN -1}{2}c\sum_{i\neq j}^{M}\frac{P'(h_{i})}{h_{i}-h_{j}}+R.
\label{eqn:bCofh}
\end{eqnarray}
The $P$ and $Q$ in Eqs.~(\ref{eqn:htham}) and (\ref{eqn:bCofh})
are a fourth- and a second-degree polynomial, respectively:
\begin{subequations}
\label{eqns:defpol}
\begin{eqnarray}
P(h)&=&a_{4}h^{4}+a_{3}h^{3}+a_{2}h^{2}+a_{1}h+a_{0},
\label{eqn:defP}\\
Q(h)&=&b_{2}h^{2}+b_{1}h+b_{0}.
\label{eqn:defQ}
\end{eqnarray}
\end{subequations}
Thus, there are 10 parameters $a_{n}$, $b_{n}$, $c$, $R$, which
characterize the quasi-solvable Hamiltonian (\ref{eqn:htham}).
One can prove the quasi-solvability of the operator (\ref{eqn:htham})
by the convertibility of it into the form (\ref{eqn:sgham}).
The function $h(q)$, which determines the change of variables,
is given by a solution of the differential equation,
\begin{eqnarray}
h'(q)^{2}=2P\bigl(h(q)\bigr).
\label{eqn:dhofq}
\end{eqnarray}
One may notice that the resultant Eqs.~(\ref{eqn:htham})--%
(\ref{eqn:dhofq}) have resemblance to those of the one-body
quasi-solvable models constructed from $\Lsl (2)$
generators~\cite{Turbi1,LoKaOl3,LoKaOl4}, or equivalently, the
type A $\cN$-fold supersymmetric models~\cite{ANST1,ANST2,Tanak2}.
Indeed, we can easily see that the above results reduce to the one-body
$\Lsl (2)$ quasi-solvable and type A $\cN$-fold supersymmetric
models if we set $M=1$, where the double summation is understood as
$\sum_{i\neq j}^{1}\equiv 0$. This is consistent with the fact
that in the case of $M=1$ the above procedure is essentially
equivalent to that in the $\Lsl (2)$ construction of type A
$\cN$-fold supersymmetry~\cite{ANST1,Tanak2}.
Under the above conditions
(\ref{eqn:htham})--(\ref{eqn:dhofq}) satisfied, the original
Hamiltonian becomes the following Schr\"{o}dinger type,
\begin{eqnarray}
H_{\cN}=-\frac{1}{2}\sum_{i=1}^{M}\frac{\partial^{2}}%
{\partial q_{i}^{2}}+\frac{1}{2}\sum_{i=1}^{M}\left[\left(
\frac{\partial\cW(q)}{\partial q_{i}}\right)^{2}
-\frac{\partial^{2}\cW(q)}{\partial q_{i}^{2}}\right]
-\bC\bigl(\sigma(h)\bigr),
\label{eqn:orham}
\end{eqnarray}
and the \textit{superpotential} $\cW(q)$ is given by,
\begin{eqnarray}
\cW(q)=-\sum_{i=1}^{M}\Int dh_{i}\frac{Q(h_{i})}{2P(h_{i})}
+\frac{\cN -1+(M-1)c}{2}\sum_{i=1}^{M}\ln\left| h'_{i}\right|
-c\sum_{i<j}^{M}\ln\left| h_{i}-h_{j}\right|.
\label{eqn:sppot}
\end{eqnarray}
It is evident by the construction that the solvable wave functions
$\psi(q)$ of the Hamiltonian (\ref{eqn:orham}) take the following form,
\begin{eqnarray}
\psi(q)=\tilde{\psi}(q)\, e^{-\cW(q)},\quad \tilde{\psi}(q)
\in\tilde{\cV}_{\cN}.
\label{eqn:svwav}
\end{eqnarray}
The Hamiltonian (\ref{eqn:orham}) with Eqs.~(\ref{eqn:bCofh}) and
(\ref{eqn:sppot}) is the most general quasi-solvable many-body
systems with two-body interactions which can be constructed from
the $\Lsl (M+1)$ generators (\ref{eqns:gensl}).

Before investigating what kind of particular models emerges from
the general Hamiltonian (\ref{eqn:orham}), we will refer to
an interesting feature of the result. If the algebraic Hamiltonian
(\ref{eqn:gsham}) does not contain any \textit{raising} operator
$E_{i0}$, it preserves the vector space $\tilde{\cV}_{\cN}$ for
arbitrary $\cN$ and becomes not only a quasi-solvable but also a
solvable model~\cite{Turbi2,Ushve}. In this case,
it turns out that $\bC(\sigma)=C$, one of the constants involved in
Eq.~(\ref{eqn:gsham}), and thus the original Hamiltonian
(\ref{eqn:orham}) becomes supersymmetric~\cite{Witte1,Witte2}.
A system is always quasi-solvable if it is supersymmetric,
since the ground state is always solvable.
From the above result, we can conclude that a system is always
supersymmetric if it is solvable and all its states have the form
(\ref{eqn:svwav}).

\section{\label{sec:class}Classification of the Models}

It was shown that the one-body $\Lsl (2)$ quasi-solvable models can
be classified using the shape invariance of the Hamiltonian
under the action of $GL(2,\bbR)$ of linear fractional
transformations~\cite{LoKaOl3,LoKaOl4}. We can see that the many-body
Hamiltonian (\ref{eqn:htham}) also has the same property of shape
invariance. The linear fractional transformation of $h_{i}$ is
introduced by,
\begin{eqnarray}
h_{i}\mapsto\hat{h}_{i}=\frac{\alpha h_{i}+\beta}{\gamma h_{i}+\delta}
\quad (\Delta\equiv\alpha\delta -\beta\gamma\neq 0).
\label{eqn:frtsf}
\end{eqnarray}
Then, it turns out that the Hamiltonian (\ref{eqn:htham}) is
shape invariant under the following transformation induced by
Eq.~(\ref{eqn:frtsf}),
\begin{eqnarray}
\tilde{H}_{\cN}(h)\mapsto\widehat{\tilde{H}}_{\cN}(h)=
\prod_{i=1}^{M}(\gamma h_{i}+\delta)^{\cN -1}\,\tilde{H}_{\cN}
(\hat{h})\prod_{i=1}^{M}(\gamma h_{i}+\delta)^{-(\cN -1)},
\label{eqn:tfham}
\end{eqnarray}
where the polynomials $P(h)$ and $Q(h)$ in the $\tilde{H}_{\cN}(h)$
are transformed according to,
\begin{subequations}
\label{eqns:tfpol}
\begin{eqnarray}
P(h)\mapsto\hat{P}(h)=\Delta^{-2}(\gamma h+\delta)^{4}P(\hat{h}),
\label{eqn:tfofP}\\
Q(h)\mapsto\hat{Q}(h)=\Delta^{-1}(\gamma h+\delta)^{2}Q(\hat{h}).
\label{eqn:tfofQ}
\end{eqnarray}
\end{subequations}
For a given $P(h)$, the function $h(q)$ is determined by
Eq.~(\ref{eqn:dhofq}) and a particular model is obtained by
substituting this $h(q)$ for Eqs.~(\ref{eqn:bCofh}),
(\ref{eqn:orham}) and (\ref{eqn:sppot}). Under the transformation
(\ref{eqn:tfofP}) of $GL(2,\bbR)$, every real quartic polynomial
$P(h)$ is equivalent to one of the following eight forms:
\begin{eqnarray*}
&&\text{1). }\frac{1}{2},\quad\text{2). }2h,\quad\text{3). }2\nu h^{2},
\quad\text{4). }2\nu (h^{2}-1),\\
&&\text{5). }2\nu (h^{2}+1),\quad\text{6). }\frac{\nu}{2}(h^{2}+1)^{2},\\
&&\text{7). }2h^{3}-\frac{g_{2}}{2}h-\frac{g_{3}}{2},\quad\text{8). }
\frac{\nu}{2}(h^{2}+1)\bigl[ (1-k^{2})h^{2}+1\bigr].
\end{eqnarray*}
where $\nu$, $k$, $g_{2}$ and $g_{3}$ are all real numbers satisfying
$\nu\neq 0$, $0<k<1$ and $g_{2}^{3}-27g_{3}^{2}\neq 0$.
Thus, the quasi-solvable models (\ref{eqn:orham}) can be classified
into the above eight cases.\\

\noindent
\textit{Case} 1). $h(q)=q$:

This leads to the rational $A$ type Inozemtsev
model~\cite{Inoze1,InMe1,Inoze3}. Inozemtsev models are known as
a family of deformed CS models which preserve the classical
integrability.
The main difference between quantum and classical case is that the
quantum quasi-solvability holds only for \textit{quantized} values
of the parameter, say, for integer $\cN$, while the classical
integrability holds for continuous values. This is one of the common
features that the quantum quasi-solvable models share.\\

\noindent
\textit{Case} 2). $h(q)=q^{2}$:

This leads to the rational $BC$ type Inozemtsev model.
The quasi-exactly solvable model reported in Ref.~\cite{HoSh1} is
just this case.\\

\noindent
\textit{Case} 3). $h(q)=e^{2\rnu q}$:

This leads to the hyperbolic ($\nu >0$) and
trigonometric ($\nu <0$) $A$ type Inozemtsev model.\\

\noindent
\textit{Case} 4). $h(q)=\cosh 2\rnu q$:

This leads to the hyperbolic ($\nu >0$) and
trigonometric ($\nu <0$) $BC$ type Inozemtsev model.
The quasi-solvability of the special cases of the above four
were recently shown in Ref.~\cite{SaTa1} by an ansatz method.\\

\noindent
\textit{Case} 5). $h(q)=\sinh 2\rnu q$:

This leads to a hyperbolic model being neither the Inozemtsev
nor the Olshanetsky--Perelomov type~\cite{OlPe2}.
The paper Ref.~\cite{UlLoRo1} covers most of the above five models.\\

\noindent
\textit{Case} 6). $h(q)=\tan\rnu q$:

This leads to a trigonometric model being neither the Inozemtsev nor 
the Olshanetsky--Perelomov type.\\

\noindent
\textit{Case} 7). $h(q)=\wp(q;g_{2},g_{3})$:

This case includes the elliptic $BC$ type Inozemtsev model and the
twisted CS models~\cite{HoPh1,HoPh2,BoSa1}. The elliptic model in
Ref.~\cite{UlLoRo2} may be also included in this case.\\

\noindent
\textit{Case} 8). $h(q)=\text{sn }(\rnu q|k)/\,\text{cn }(\rnu q|k)$:

This leads to an elliptic model being neither the Inozemtsev nor
the Olshanetsky--Perelomov type.\\

The two-body potentials in all the cases have singularities at
$q_{i}=q_{j}\ (i\ne j)$. Thus, each of the models is naturally defined
on a Weyl chamber if the potential is non-periodic or on a Weyl alcove
if the potential is periodic~\cite{OlPe2}. Cases 1--5 with $\nu >0$
and Case 6 with $\nu <0$ correspond to the former while the others
to the latter. In the latter case, a system can be quasi-exactly
solvable unless a pole of a one-body potential in the system exists
and is in the Weyl alcove.
On the other hand, quasi-exact solvability in the former case
depends mainly on the asymptotic behavior of Eq.~(\ref{eqn:svwav}) at
$|q_{i}|\to\infty$. Since this behavior is in general not dominated by
the two-body term in the r.h.s. of Eq.~(\ref{eqn:sppot}), most
of the results on the normalizability of the one-body $\Lsl (2)$
quasi-solvable models in Ref.~\cite{LoKaOl3} may also hold for our
models.

More details on the results presented here and further development
will be reported in the near future~\cite{Tanak3}.
\begin{acknowledgments}
We would like to thank H.~Aoyama, N.~Nakayama, R.~Sasaki, M.~Sato,
 K.~Takasaki and S.~Yamaguchi for useful discussions.
We would also like to thank A.~Gonz\'{a}lez-L\'{o}pez for crucial
comments.
This work was supported in part by a JSPS research fellowship.
\end{acknowledgments}



\end{document}